\documentclass[useAMS,usenatbib, useAMSmath]{mn2e}
\usepackage{longtable}
\usepackage[dvips]{graphics}
\usepackage[pdftex]{graphicx}
\usepackage{amssymb}
\usepackage{natbib}
\title[Optical Polarimetric Observations of PKS 2155-304]{Particle
  Acceleration and Magnetic Field Structure in PKS 2155-304: Optical
  Polarimetric Observations}
\author[U. Barres de Almeida et al.]{U. Barres de Almeida$^{1}$\thanks{E-mail:
u.b.almeida@durham.ac.uk}, M.J. Ward$^{1}$, T.P. Dominici$^{2}$,  Z. Abraham$^{3}$, G.A.P. Franco$^{4}$,\newauthor M.K. Daniel$^{1}$, P.M. Chadwick$^{1}$ and C. Boisson.$^{5}$\\
$^{1}$Department of Physics, University of Durham, South Road, DH1 3LE, England.\\
$^{2}$Laborat\'{o}rio Nacional de Astrof\'{i}sica, Rua Estados Unidos
154, 37504-364, Itajub\'{a}, Brasil.\\
$^{3}$Departamento de Astronomia, Universidade de S\~{a}o Paulo, Rua do
Mat\~{a}o 1226,05508-090, S\~{a}o Paulo, Brasil.\\
$^{4}$Departamento de F\'{i}sica, ICEx - UFMG, Caixa Postal 702,
30123-970, Belo Horizonte, Brasil.\\
$^{5}$Observatoire de Paris-Meudon, 5 Place Jules Janssen, 92190,
Meudon, France.\\
}
\begin{document}

\date{Accepted year Month day. Received year Month day; in original form year Month day}

\pagerange{\pageref{firstpage}--\pageref{lastpage}} \pubyear{2010}

\maketitle

\label{firstpage}

\begin{abstract}
In this paper we present multiband optical polarimetric observations
of the VHE blazar PKS 2155-304 made simultaneously with a H.E.S.S./Fermi
high-energy campaign in 2008, when the source was found to be in a low
state. The intense daily coverage of the dataset allowed us to study
in detail the temporal evolution of the emission and we
found that the particle acceleration timescales are decoupled from
the changes in the polarimetric properties of the source. We present a
model in which the optical polarimetric emission originates at the polarised
mm-wave core and propose an explanation for the lack of correlation
between the photometric and polarimetric fluxes. The optical emission 
is consistent with an inhomogeneous synchrotron source in which the
large scale field is locally organised by a shock in which particle
acceleration takes place. Finally, we use these optical polarimetric
observations of PKS 2155-304 at a low state to propose an origin for the 
quiescent gamma-ray flux of the object, in an attempt to provide
clues for the source of its recently established persistent TeV emission.   
\end{abstract}

\begin{keywords}
polarization; galaxies: jets; BL Lacertae objects: PKS 2155-304.
\end{keywords}

\section{Introduction}

\begin{figure*}
\begin{center}
\rotatebox{-90}{\includegraphics[width=9.7cm, height=15.5cm]{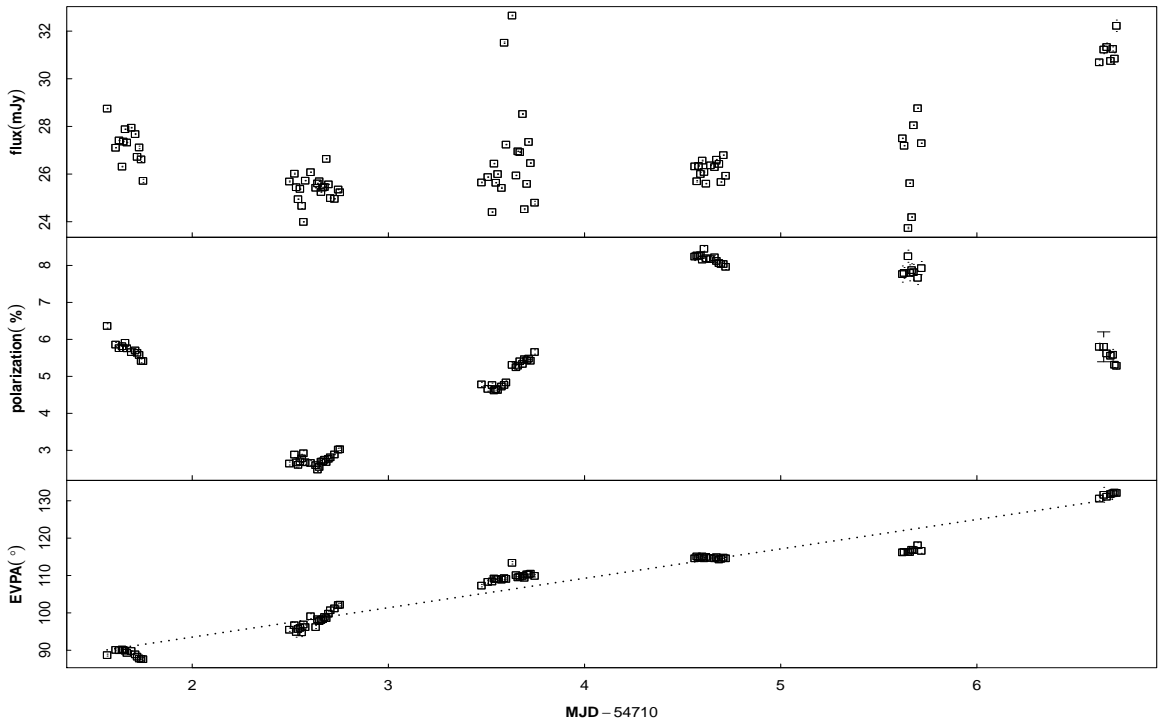}}
 \caption{R-band optical light-curve for PKS 2155-304 from 01
   to 07 September 2008, showing the total flux variability
   (\textit{upper panel}), fractional polarisation degree (\textit{middle
     panel}) and EVPA rotation (\textit{lower panel}). Each data point
   corresponds to an integration time of about 6 min. The flux points
   and polarisation degree are not corrected for the host galaxy
   contribution. Error bars are of the order of the size of the points.}
\end{center}
\label{pollc}
\end{figure*}

Strong and variable linear polarisation is a defining property of
BL Lacs, which unavoidably associates the observed emission with 
beamed synchrotron radiation from a relativistic jet viewed
face-on. The polarimetric properties of these sources have been thoroughly
studied at radio and optical wavelengths for over three decades,
displaying rich phenomenology (e.g. \citealt{b0} and \citealt{saik}). The
inner jets of active galaxies present a morphology characterised by a
stationary region of enhanced brightness, the ``core'', and other
luminous components (``knots'') moving at relativistic speeds
and thought to be associated with the propagation of shock
perturbations. These structures tend to be highly
polarised in radio and are widely recognised as potential sites of
particle acceleration and as responsible for the observed flux
variability \citep{hughes}. Core brightening and the
appearance of new knots in the jet have been linked to flaring
activity extending to gamma-ray energies \citep{jorstad01}.      

These shocked regions are persistent and bright structures in the X-ray
images of mis-aligned pc-scale jets, and have also been invoked to explain
the quiescent level of emission observed from some blazars at these
energies (e.g. \citealt{b5}). In blazars, these knots are not
resolved due to the close alignment of the jet to the line of sight,
but their existence is implied by analogy with similar objects, such as
M 87 \citep{marshall}. Despite the fact that in mis-aligned objects
the knots are responsible for only a fraction of the flux emitted from
the source, the geometrical alignment in blazars provides the
required amount of Doppler boosting necessary to produce the low-level of
continuous emission observed \citep{b5}. 

\cite{kat} have recently
modeled the full spectral energy distribution (SED) of PKS 2155-304
and identified the presence of such a background synchrotron
component, which is associated with an extended jet component
(presumably the integrated contribution from all X-ray-bright knots
along the jet) capable of sustaining the low state X-ray flux observed
from the source.   
At very-high energies (VHE), consistent detection of the BL Lac PKS
2155-304 by the H.E.S.S. telescopes has also established a low-level
of $\gamma$-ray emission for this object which is stable over several
years and strongly constrains the blazar's VHE quiescent state
\citep{prep}, but whose origin is still unknown. 

The low state of PKS 2155-304 ($z = 0.116$) has been studied recently
in a multiwavelength
campaign by H.E.S.S. and the LAT instrument onboard Fermi \citep{b1c}, with
which the observations presented here are simultaneous. The
time-averaged SED of the source was modeled as a single-zone
synchrotron-self Compton (SSC) 
process which fitted the entire profile. The derived relations between the
optical and VHE fluxes suggest that the former provides the
target photons for the inverse-Compton (IC) emission, but a detailed
study of the energetics of the model shows that a single-zone
description cannot accommodate the entire multiband temporal behaviour of the 
light-curve. Part of the aim of this paper is to exploit the
contemporaneity of our data to propose an explanation for the lack of 
temporal correlation observed between the optical and the high-energy
components of the SED, and to give further support to the proposal by 
\cite{b1c} that a multi-zone model is necessary to describe the
quiescent state emission of this BL Lac.
         
PKS 2155-304 has been the target of several optical polarimetric campaigns
which have probed its long and short term behaviour. The polarisation
degree is observed to assume typical values between $\sim$ 3-7\%, with
significant variability registered down to sub-hour timescales
(e.g. \citealt{micro}). The polarisation vector shows evidence of a
preferential direction within the range 100-140$^\circ$
(\cite{tommasi} and references therein). Frequency dependent polarisation (FDP)
has also been detected on several occasions 
\citep{b22a, b22c, allen} and determined to be
intrinsic to the synchrotron source. In radio, the parsec-scale jet
of PKS 2155-304 was imaged twice at 15 GHz by \cite{b20} and
\cite{b21}. A single jet component is resolved downstream from the
radio core, moving with a derived bulk Lorentz
factor $\Gamma \sim$ 3. Polarised radio flux was detected in those
images coming from the core component alone, and the polarisation
vector ($131^\circ$) was seen to be closely aligned with the
jet-projected position angle (P.A. $\sim$ 140-160$^\circ$). In the
optically thin regime this is evidence for the presence of a dominant 
magnetic field component transverse to the flow. The polarisation
degree of the core exhibited a spatial gradient between 3-8\% that
increased in the upstream direction. 

The existence of a preferred position angle in optical similar
to that of the mm-wave core favours the presence of a dominant or
large scale component with a regular magnetic field which is associated
with both emissions. Furthermore, similar values of the
polarisation degree seen in both bands and the lack of polarised
emission from other parts of the jet in the VLBI images suggests
the unresolved polarised optical emission originates in the pc-scale
radio core. This hypothesis will be adopted here, motivated as well by
recent studies which used VLBI maps to compare the optical
polarisation properties of the jet with the radio images, and associated
the variable emission with the position of the 43 GHz core \citep{b15,
jorstad07, gabuzda}.

\section{Description of the Observations}

The optical polarimetric campaign was conducted in early September 2008, 
between MJD 54710-54716. Observations
were made with the 1.6 m Perkin-Elmer telescope at Pico dos Dias
Observatory of the National Astrophysics Laboratory (OPD/LNA,
Brazil), using the imaging polarimeter IAGPOL in linear
polarisation mode. Multi-band images in the V, R and I filters were taken on
all nights but the last of the campaign. The configuration of the polarimeter 
provides simultaneous measurements of the ordinary and extra-ordinary
rays, allowing us to perform observations under non-ideal atmospheric
conditions, since any atmospheric contributions will affect both rays equally;
additionally, any sky contribution is expected to cancel out in the
process. Photometric flux measurements are obtained simultaneously
with the polarimetric ones. Standard polarisation
stars from \cite{b22b} and \cite{b22rp} were used for calibration. Single
polarisation images were integrated from 8 $\times$ 45 s exposures,
each at a different position of the polarimetric wheel. A precision of
better than 1\% in the polarisation degree was achieved. The temporal
resolution of consecutive measurements in the R band was of the order
of 8-10 min, whereas V and I images were taken at the beginning and end
of each night to monitor the spectral evolution of the source. Data
reduction was made with a specially developed analysis package for
LNA polarimetric data, PCCDPACK \citep{b19}.

\section{Discussion of the Optical Polarimetric Data}

\begin{figure}
\begin{center}
\rotatebox{-90}{\resizebox{6.5cm}{8.5cm}{\includegraphics{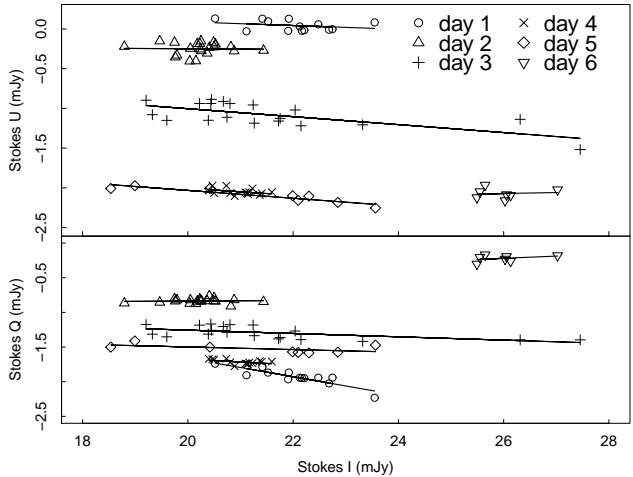}}}
 \caption{Polarised versus total flux relation for the six nights of
   the campaign. The straight lines are fits to the data used to
   derive the polarisation properties of the variable component.} 
\end{center}
\label{microvar}
\end{figure}

\begin{table}
 \caption{Polarisation parameters of variable component.}
 \label{symbols}
  \begin{center}
  \begin{tabular}{@{}lcccccc}
  \hline
 ~~~MJD~~~~   & ~~~~~~$p_{\rm{var}}$~~~~~ & ~~~~~~$\theta_{\rm{var}}$~~~~~~&~~~~~$I_{\rm{var}}$~~~~~ \\
        & ~~~~~(\%)~~~~~ & ~~~~~($^\circ$)~~~~~&~~~~~(mJy)~~~~~ \\
  \hline
  ~~~54711 & $12.5 \pm 1.3$ & $84.9 \pm 5.6$  & $2.3 \pm 0.6$ \\
  ~~~54712 & ~$1.0 \pm 0.6$ & $70.0 \pm 12.0$ & $2.0 \pm 0.2$\\
  ~~~54713 & ~$5.6 \pm 1.4$ & $102.2 \pm 7.0$ & $3.8 \pm 0.6$\\
  ~~~54714 & ~$7.5 \pm 1.4$ & $120.1 \pm 6.4$ & $1.8 \pm 0.8$\\
  ~~~54715 & ~$6.8 \pm 1.3$ & $123.6 \pm 6.2$ & $5.8 \pm 0.8$\\
  ~~~54716 & ~$3.4 \pm 1.9$ & $125.4 \pm 6.5$ & $7.5 \pm 1.0$\\
  \hline
 \end{tabular}
 \end{center}
\medskip
% $^*$The values of the polarisation angle carry an intrinsic
%ambiguity of 180$^\circ$.
\end{table}

Figure 1 shows the R band light-curve for the total flux, polarisation
fraction and electric vector position angle (EVPA) for all six
nights of the optical campaign. The data presented in this figure (see
also Table 2 at the end of the paper) represents the directly observed
quantities, not corrected for the unpolarised contribution of the
stellar continuum. For the remainder of the analysis, flux estimates
as quoted in \cite{dominici} were used to subtract the host galaxy
contribution to the total emission. 

The source was observed for three to six hours during each night with
a minimum temporal resolution in the R band of
$\sim$ 10 min, resulting in a week of well sampled intranight light-curves. 
The overall flux behaviour is qualitatively distinct from the changes in the 
polarisation properties of the emission, as noted before by
\cite{cour} and \cite{tommasi} for this same object. Flux variability is
dominated by intranight activity, superimposed on a baseline level which
increases towards the end of the campaign and is in agreement with the
measurements from the ATOM telescope presented in \cite{b1c}. A Lomb-Scargle
power spectrum analysis \citep{b22} reveals that the flux
microvariability is describable as random fluctuations, 
with minimum variability timescales $<$ 1 hr, limited
by the sampling of the lightcurve. 

Although presenting some intranight activity, the temporal behaviour of the
polarised flux was dominated by inter-night variations with larger
relative amplitude than those of the unpolarised flux, varying by a
factor of 3 during the campaign. The host-corrected polarisation degree
varied smoothly between 3-11\% along the six nights of
observations, within the range typically registered for the 
source and similar to those seen for the radio core. A very similar 
``oscillatory'' behaviour for the polarisation fraction
can be seen in the optical lightcurves of \cite{cour}, but the
behaviour of the polarisation vector is very distinct at both
epochs. 

Visual inspection of the light-curves shows that the total 
photometric variability cannot be explained by variations in
the polarised flux alone. Subtraction of the polarised flux from the
photometric light-curves leave residual variability both in the
intranight and the long-term flux variations. Conversely, 
dilution of a constant polarised component on a variable, unpolarised 
background cannot account for the observed behaviour of the
polarisation degree, which changes in an uncorrelated fashion with
respect to the total flux. Throughout our observations the EVPA
underwent a quasi-linear counter-clockwise rotation of about
$40^\circ$, at a rate of $\approx$ 7$^\circ$ per day. The lack of
correlation between the smooth, long-term evolution of the
polarisation parameters and the flux behaviour is a common property of
BL Lacs \citep{qian} in optical and must be explained.

\subsection{Modelling of the Polarised Emission}

\begin{figure}
\begin{center}
\rotatebox{-90}{\resizebox{7.8cm}{8.6cm}{\includegraphics{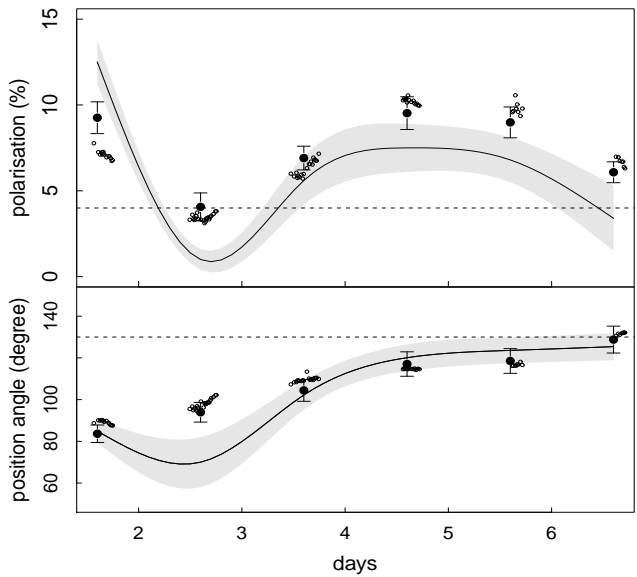}}}
 \caption{Results of the two-component synchrotron model fit to 
   the optical polarimetric data of PKS 2155-304 (small dots). Large
   filled circles are the resulting model values for each night,
   obtained as the result of the superposition of a constant (dashed
   lines) and variable polarised component (smooth solid line) using
   the law of addition of polarised sources (Equations (1) and
   (2)). The smooth curves are a spline interpolation to the model
   values presented in Table 1 and show the temporal evolution of the
   variable component. Grey shades represent the confidence intervals
   as quoted in Table 1. The values for the constant component
   presented in the plots are $p_{\rm{cons}}= 4 \pm 1\% $ and
   $\theta_{\rm{var}}=130^\circ \pm  10^\circ$ for a ratio of fluxes 
   $I_{\rm{var}}/I_{\rm{cons}} < 1$. Notice in the bottom plot the
 rotation of the position angle of the model variable component
 towards a gradual alignment with the direction of the constant component.} 
\end{center}
%\label{poldegfit}
\end{figure}

If the observed variability within a given time interval is due to a 
\textit{single variable} component with \textit{constant} polarisation 
properties, a linear relationship exists between the absolute Stokes 
parameters Q and U and the total flux, as described in
\cite{b11}. Figure 2 shows that although this relation is not obeyed
by the entire dataset collectively, intranight measurements taken individually
clearly follow a linear trend. This suggests that the flux microvariability
could be either the result of a single variable component whose
Stokes parameters evolve on longer timescales than those of the
intranight monitoring, or represent the manifestation of several different
components with different polarisation properties dominating the
emission on each night. The smoothness of the temporal evolution of
the polarisation parameters seen in Figure 1 seems nevertheless to
disfavour the presence of a great number of components, each active at 
different times. In particular, the fact that the 
polarisation properties of PKS 2155-304 change more slowly than the
total flux argues against the polarised flux being the resultant 
of the contribution of a large number of independent components.   

From the fits to each set of intranight measurements presented in
Figure 2, relative Stokes parameters can be directly determined as the
slopes of the lines and these are used to model the polarisation properties
for the variable component, $p_{\rm{var}}$ and $\theta_{\rm{var}}$,
presented in Table 1. Although an appropriate physical description for this
variable component has not yet been given, its existence is directly
implied from the analysis shown in Figure 2 and the observational
motivation behind its identification is to single
out a particular region of the source through which \textit{all}
photo-polarimetric variability might be explained. The presence of
this hypothetical component will now be tested by means of a formal
modeling of the emission. 

The polarisation degree for the putative variable component as
determined from Figure 2
varied in the range 1-13\% during the campaign, reaching a
minimum on the second night, when its intrinsic polarisation almost
disappeared. Although the temporal evolution of $p_{\rm{var}}$ and 
$\theta_{\rm{var}}$ broadly follows
the same trend of the integrated source's polarisation, it does not
match exactly the observed parameters in Figure 1. This mis-match in
the polarisation properties suggests the presence of another polarised 
component by which this variable emission is ``diluted''. This is
particularly evident from the fact that the EVPA derived for the
variable component does not agree with the values measured for the
source's polarisation angle at all epochs.

The interplay between a \textit{constant polarised} component,
associated with the underlying jet, and a \textit{variable} one due to
the propagation of a shock has been proposed by a number of
authors to explain a variety of variability behaviours in blazars 
(e.g., \citealt{holmes}, \citealt{qian93}, \citealt{brindle}). Due to
the lack of strictly simultaneous multi-band data, we will attempt a
model only to the monocromatic R-band observations. The resulting
polarisation properties of the superposition of two optically thin synchrotron
components are given by the following equations \citep{holmes}:

 \begin{eqnarray}
p^2 =\frac{p_{\rm{cons}}^2+p_{\rm{var}}^2~I_{\rm{v/c}}^2+2~p_{\rm{cons}}~p_{\rm{var}}~I_{\rm{v/c}}\cos2\xi}{(1+I_{\rm{v/c}})^2}\\
\tan2\theta=\frac{p_{\rm{cons}}~\sin2\theta_{\rm{cons}}+p_{\rm{var}}~I_{\rm{v/c}}~\sin2\theta_{\rm{var}}}
{p_{\rm{cons}}~\cos2\theta_{\rm{cons}}+p_{\rm{var}}~I_{\rm{v/c}}~\cos2\theta_{\rm{var}}}
%\label{eq02}
\end{eqnarray}\\
where $\xi = \theta_{\rm{cons}}-\theta_{\rm{var}}$ and $I_{{\rm{v/c}}}$
is the ratio of fluxes of the variable and constant components. 

In order to determine the values for the parameters of the constant
component and the ratio of fluxes $I_{{\rm{v/c}}}$, we followed a fitting
procedure similar to that of \cite{qian93}. Given the uncertainties in 
$p_{\rm{var}}$ and $\theta_{\rm{var}}$, and the complex trigonometric
relations in Equations (1) and (2) which prevent a straightforward
analytical solution, the fitting process had to be done iteratively. 
We chose to start from the second night, where the contribution of the
variable component was likely to be smallest and used Equations (1)
and (2) to find the best fitting values for the parameters $p$ and
$\theta$ for both the variable and constant components. This was done
automatically by searching the entire parameter space and 
minimising the model residuals. This procedure gave us an estimate for
the flux level of the constant underlying jet component,
$I_{\rm{cons}} \approx$ 20
mJy. Its polarisation degree $p_{\rm{cons}}$ was also estimated from the
same dataset to be $\sim 3\%$. The best-fit value for
$\theta_{\rm{cons}}$ corresponding to these polarisation parameters
was of $\approx 120^\circ$. Analysis then proceeded by 
applying this set of values as a starting
point for the fit to each individual night. The parameters of the constant
component were allowed to vary within the same error range of
those of the variable component as quoted in Table 1, since they
indicate the limiting accuracy of the model fitting. We looked for
the values of the variable component on each night which minimise the
residuals while keeping our pre-determined bounds
for the constant component, i.e. $I_{\rm{cons}} \approx$ 19-21 mJy, 
$p_{\rm{cons}} \approx$ 1-5 $\%$ and  $\theta_{\rm{var}} \approx$
110-140$^\circ$. An acceptable solution was found for every night, and
the residuals were kept below 10\% for each dataset.

A good indication of the appropriateness of this
model in describing the entire dataset is that a
reasonable fit was obtained for each night without the need for the parameters of the
constant component to depart from the boundaries mentioned above. Such
boundaries are regarded as indicating the range of accuracy within which
the component's parameters can be regarded as ``constant'', since they
reflect the intrinsic uncertainty of the fitting as defined by the
errors in the parameters of Table 1. Final confidence intervals for the polarisation
parameters of the constant component were estimated from the range of 
night-to-night variations in its best-fit parameters, and are given by
$p_{\rm{cons}} = 4 \pm 1\%$ and $\theta_{\rm{cons}} = 130^\circ \pm
10^\circ$. They are therefore compatible with a set of constant
parameters throughout the campaing within the observational errors.
This best-fit model is shown in Figure 3. For all
nights we had $I_{\rm{v/c}} < 1$, indicating that the background
component dominates the photometric flux emission. The values of $I_{\rm{var}}$
derived for each individual night are presented in Table 1,
corresponding to 15-45\% $I_{\rm{cons}}$.

The derived parameters for the constant component are found to match
the regular values of the polarisation 
compiled by \cite{tommasi} for PKS 2155-304, suggesting its
association with a persistent optical jet component. The degree
of polarisation $p_{\rm{cons}}$ is also similar to the minimum values
measured for this source at 43 GHz and in historical optical data, and 
the corresponding position angle is well aligned with the radio-core
EVPA as determined by \cite{b21}. This coincidence also attests to the
presence of a field component in the jet which is common both to the
radio and optical wavelengths and persistent in time, and whose direction is
transverse to the flow, as expected from a shock-compressed tangled
field. 

From the second night of the campaign onwards, the position angle of
the variable component rotated continuously from $70^\circ$ (i.e. approximately 
$90^\circ$ mis-aligned with the jet-projected P.A.) to $\approx
120^\circ$, in close alignment with the direction of the persistent
jet component. The rotation of $\theta_{\rm{var}}$ could be
interpreted as the gradual alignment of the field of a new ``blob'' of 
material, as it encounters a shock in the core that re-organises
its field. The maximum value observed for the source's polarisation
degree coincides with the epochs of greatest alignment between the two
fields, and the start of the rotation in $\theta_{\rm{var}}$ marks the
onset of the increase on the baseline photometric flux seen towards
the end of the campaign. Such a scenario, where both optical position angles
$\theta_{\rm{var}}$ and $\theta_{\rm{cons}}$ tend to align with the
direction of the radio EVPA when the
observed polarisation is high, was considered before by \cite{valb} for
the quasar 3C 273 during a radio-to-optical flare. In such a scenario
a correlation is expected between the optical and polarised fluxes
which is marginally observed in our dataset, and more observations at
more active source states are necessary to better establish the validity of
the correlation for this object.

\subsection{Origin of the Flux Variability}

As noted before, the observed flux variability happens on two
different timescales, its amplitude being dominated by intranight
variability, superimposed on a background level that steadily 
increases towards the end of the campaign, and which we have
associated to the evolution of the variable (or shocked) component in the model
of the previous section. 

\subsubsection{Microvariability}
To try to identify the physical origin of
these variations and in particular the nature of the flux
microvariability, we observe that the intranight flux changes
were accompanied by changes in the spectral index. The source
presented colour variations both in intranight timescales and in the
nightly averages. The intranight $(V-I)$ colours varied in the range
0.12 -- 0.27, with greatest amplitude in the third night of the campaign,
when variability was the greatest. Colour variations can be
linked to radiative cooling of electrons in a magnetised plasma,
implying synchrotron lifetimes of the order of the intraday timescales
of a few hours. The synchrotron lifetime in the observer's frame,
written in terms of the observed photon frequency in units of GHz,
$\nu_{GHz}$, is given by \citep{pac}:

\begin{equation}
t_{sync} \approx 1.1\times10^{4}\left(\frac{1+z}{\delta \nu_{GHz}
  B_{G}^{3}}\right)^{1/2} \rm{hours}
\label{eq02}
\end{equation}

For $t_{\rm{sync}}$ equal to the timescales of intranight variations in the
R band, and using typical Doppler factors for PKS 2155-304
of about $\delta \sim 30$ (e.g. as for the compact components in \citealt{kat})
we obtain a magnetic field $B \lesssim 0.5$ G for the variable
component. 

The fact that we see such changes in colour 
simultaneously with the intranight variations, suggests they can
be taken as a direct signature of particle acceleration and cooling at
the source, with  $t_{\rm{acc}} < t_{\rm{sync}}$. An upper limit to the size
of the acceleration region can then be set to $r_{\rm{s}} < \delta
t_{\rm{sync}}c/(1+z) \approx 10^{16} \rm{cm} \sim 5 \times 10^{-3} \rm{pc}$. 
This limit is in accordance with predictions for the thickness
of shocks given by \cite{shock} and points to an origin for the flux 
microvariability as the result of particle acceleration taking place
at a shock front, with high magnetic field due to plasma compression. 
Magnetic fields of this order have also been considered by
\cite{shock} as typical estimates for the field intensity in blazar
cores, and are of the same order of magnitude of those recently found
to explain the low-activity state of Fermi-detected blazars
\citep{Fermib}. In the SED model of \cite{kat} such values for the
{\bf B}-field and Doppler factor are also associated with the 
variable shocked components, as opposed to the extended jet which had
lower values for both parameters.

Our values for $\mathbf{B}$ are in fact an order of
magnitude higher than those derived by \cite{b1c} from an SED fit to
the data of the H.E.S.S./Fermi campaign. The parameters calculated by \cite{b1c}
correspond to those obtained by \cite{kat} for an SSC description of the
steady component in the SED of PKS 2155-304. A steady jet
component with these properties, although showed to be responsible for
the persistent X-ray emission, cannot explain the rapid flux and
spectral variability that we observe, nor can it properly account
for all aspects of the IC emission, as discussed by \cite{b1c}. \cite{kat}
noticed that the IC emission of their background component was
in fact negligible in the TeV range at the time of their
observations. This suggests that whereas the
bulk of the optical-to-X-ray synchrotron emission could be
related to the extended jet component (in fact $I_{\rm{c}}$
contributes the most to the optical flux in our model),
the highly polarised and variable optical emission (and possibly also
the VHE inverse-Compton flux) is most likely associated to the more 
energetic region, with
stronger {\bf B}-fields and $\delta$, which we identified with
the variable component modeled in the previous section. 
  
The success obtained by \cite{b1c} on explaining the broad features of the
time-averaged SED of PKS 2155-304 by a single-zone SSC
component, with parameters similar to those of the extended jet as
determined by \cite{kat}, could be attributed to the flux dominance on
the part of the extended component, which in our observations accounted
in average for 3/4 of the total optical emission. The dominance of a
single extended component to the optical synchrotron flux could also explain
why the quiescent states of BL Lacs are usually well described as a
single-zone SSC. 

The necessity of a multi-zone scenario as postulated 
by \cite{b1c} to explain the temporal behaviour of the source during
the 2008 quiescent state is nevertheless in agreement
with the view proposed here that the polarised emission
needs two dominant components to be satisfactorily explained. This scenario could also
suggest that the variable polarised emission in BL Lacs might be more clearly
linked to the IC flux and thus be a better tracer of the high-energy
behaviour than the total integrated optical flux. It also demonstrates
that optical polarimetric measurements can give crucial information
about the source structure which otherwise cannot be unambiguously
obtained, even in the context of contemporaneous multiwavelength observations.

\subsubsection{Internight Variations}

In the model presented in Section 3.1, the long-term increase
of about 5 mJy in the ``baseline'' flux level of the variable
component towards the end of the campaign was associated to a
flux increase of the variable component. The intrinsic
(host-corrected) average nightly $(V-I)$ colours for the source varied
between -0.17 to -0.01 mag, and were bluest towards the end of the
campaign, correlating with the increase observed in the baseline
photometric flux level. If we assume that the intrinsic colours observed for the
second night (when the source's flux was the lowest; $(V-I)_{\rm{cons}}$ =  -0.01) are
representative of the colours of the extended jet component, then we
can explain the changes in the average nightly colours as the
superposition of a redder, stable spectral component (due to the jet) and a bluer one,
variable at both intranight and internight timescales and due to the shock. In this
case, the changes in colour by $\Delta(V-I) = $ -0.16 mag, associated with
the brightening of the source during the last nights of the campaign
would be due to the relative increase in the flux of the variable
component, as expected from the evolution of a growing shock.

\section{Magnetic Field Structure}

Synchrotron emission from an optically thin plasma
will produce radiation that is naturally linearly polarised, with a
degree of polarisation which is dependent on the amount of ordering of
the magnetic field within the source, its spatial orientation and the
pitch-angle distribution of the radiating electrons, the latter
usually assumed to be uniform \citep{pac}
In the optically thin regime, the polarisation is a direct indicator
of the state of the magnetic field {\bf B} inside the
emission volume. If the source is
inhomogeneous its observational properties will result from
the integrated characteristics of all different emitting regions, and
will generally lead to a decrease of the net polarisation
degree while revealing any large-scale anisotropy or symmetry in the structure
of the magnetic field \citep{jones}. Wavelength or time-dependent
polarisation properties will
result from inhomogeneities and can be used to trace the
internal structure of the source. Turbulence in the flow is one
such possible source of inhomogeneities, affecting the magnetic field
structure and breaking its overall coherence beyond some 
characteristic scale $l_{B}$ \citep{jones88}. 

\subsection{Polarisation Variability}

The absence of correlation between the variations of the polarisation
degree and photometric flux and in particular the lack of
counterparts in the polarisation degree for the microvariability
suggests that the timescales of evolution of the magnetic field are
decoupled from those of particle acceleration by the shock. To
investigate the magnetic field structure in our shock-in-jet scenario, 
we follow a stochastic analysis proposed by \cite{jones}. He shows
that the spatial scale of magnetic field disorder $l_{B}$ can be
directly estimated from the intrinsic degree of polarisation of the
source $\kappa$, after correcting for the contribution of any
unpolarised emission. Here we adopt the properties of
the underlying component in the model of Section
3.2 as representative of the underlying jet parameters. We take the
internight scatter in the polarisation degree to be of the order of
the uncertainty in the parameter $p_{\rm{cons}}$, that is $\delta p
\sim 2\%$. With this, we can estimate the coherence length of the
large-scale field as being $l_{B} = (\kappa \Pi_{0}/\delta p)^{-2/3}l \sim
0.15~l$, where $\Pi_{0} =0.7$ is the polarisation fraction of a
perfectly ordered field region, and $l$ is the size of the emitting
source. If the optical emission comes from a region with size of the
order of the VLBI radio core, then $l \approx 1$ mas \citep{b21}
and $l_{B} \approx 0.3~\rm{pc}$. 

This linear scale can be compared with shocked-jet models (e.g.,
\citealt{shock}), in which the inter-day variability is associated with
the distance along the jet travelled by the relativistic shock in the
time between two extrema of the light-curve. The expression of the
distance travelled by the shock is \citep{qian,rees}:

\begin{equation}
\Delta t = \left(\frac{D(1+z)}{c\beta_{s}\delta_{s}\Gamma_{s}}\right)
%\label{eq01}
\end{equation}      

Using values derived by \cite{b20} and \cite{b21} for the
shock speed ($\beta_{s} = 1-4$), bulk Doppler factor ($\delta_{s}
\sim 30$) and Lorentz factor ($\Gamma_{s} \approx 3$), and taking
$\Delta t = 2$ days, the timescale between extrema in the polarisation
lightcurve, we obtain $D \approx 0.3$ pc for the distance traveled by the
shock. This distance, being consistent with the field turbulence scale
$l_{B}$, suggests a connection between the internight variations observed in the
polarisation degree and the spatial changes in the magnetic field,
induced by inhomogeneities in the jet. As pointed out by \cite{qian},
if these inhomogeneous structures are ``illuminated'' by the shock
through amplification of the magnetic field and increased electron
density, it will induce changes in
the integrated polarisation parameters. The timescales for these variations are
thus not necessarily associated to the fast variations in flux
observed due to particle acceleration and cooling at the shock
front. On the other hand, the increase on the total optical flux that
is seen towards the final nights of the camapign can still be
associated with these inhomogeneities since the associated changes
in the electron density will enhance the emissivity of the variable
component.
 
\cite{mckinneya, mckinneyb} have performed
general-relativistic MHD simulations of jets which show the development 
of current-driven instabilities beyond the Alfven surface ($\gtrsim 10^3$
gravitational radii, $r_{g}$). These instabilities can induce the formation of 
structures in the jet (called ``patches'' -- see Figure 2 in
\citealt{mckinneya}) characterised by an enhanced Lorentz factor and
distinct physical properties to the rest of the jet, such as magnetic
field intensity and particle density, which can drive internal
shocks. The typical sizes of these ``patches'' can be as
large as $\sim 10^3 r_{g}$, which in the case
of PKS 2155-304 is equivalent to $0.1-0.2$ pc, and thus not
very different from the estimated coherence length of the field
derived above. If such structures indeed develop in the inner
regions of AGN jets, they could provide the right scale of
inhomogeneities necessary to explain the variations on flux and 
polarimetric properties that we observe as the timescale necessary for
the shock to traverse one of these ``patches''.

As noted by \cite{tommasi}, the lack of correlation in the
photo-polarimetric properties of blazars, such as observed here, is
not easily solved by simply invoking models where the variability of
both components has origin at distinct physical regions. Stochastic
models which advocate the appearance/fading of a number of
short-lived components with independent
polarimetric properties (such as proposed by \citealt{moore}),
though able to explain the randomness of the variability as well as the lack of
photo-polarimetric correlation and the relatively low polarisation
levels observed during low states, cannot account for the
existence of a preferred position angle to the source's polarisation. 
\cite{cour} also observes that changes in the beaming factor would require a
complex geometry of the source to accomodate the poor correlation
between the total and polarised fluxes, and so an explanation relying
on aberration was also disfavoured by us.

In the picture we put forward, particle accceleration and cooling
happening at the shock front are responsible for the fast flux
variability. Variations on the polarisation degree are
associated to the propagation of this same shock through an
inhomogeneous plasma, compressing and re-ordering its otherwise tangled field
\citep{laing}. The longer timescales for the change of the polarisation
degree thus result from the shock encountering portions of
the jet which have different magnetic field properties, leading to a
changing ratio of ordered to chaotic magnetic field intensity, as
derived from the integrated source emission. It is important to stress
that this scenario can naturally explain the lack of correlation
between the photometric and polarised fluxes, whilst associating the
origin of both phenomena to the same physical region, namely an
evolving shock. If the scenario proposed by us is correct,
than polarimetric observations can serve as important diagnostics of the
structure of the magnetic field in the source, on scales that are
directly related to those of the variability of the polarised flux and
provide tighter constraints to the location and nature of the emission sites.

\begin{figure}
\begin{center}
\rotatebox{0}{\resizebox{8.5cm}{7.2cm}{\includegraphics{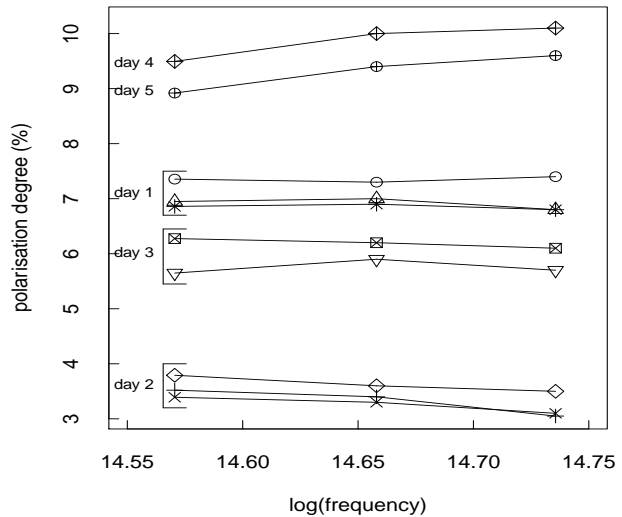}}}
 \caption{Frequency dependence of the polarisation degree. Each
   sequence of points of the same type are connected by a solid line and 
   represent quasi-simultaneous I, R and V measurements of the
   polarisation degree. The annotations to the left of the data points
   indicate the dates of the campaign corresponding to each set of measurements. 
   The vertical scale orders the measurements according to total intensity
   of polarisation and show that FDP increasing with frequency is
   present at high polarisation levels.} 
\end{center}
\label{poldegfit}
\end{figure}

\subsection{Frequency Dependent Polarisation}

Spectral dependence of the polarisation is a common feature of blazars
and its study gives information about the structure of the synchrotron
source. To search for the presence of FDP we use the I and V band
measurements taken at the beginning and end of each night, within
approximately 30 min of observations in the R-band. \cite{b6} showed
that the polarisation of radiation from a homogeneous synchrotron
source with a power-law distribution of electron energies is
frequency-independent, and so the presence of FDP is indicative of 
inhomogeneities in the particle distribution or magnetic field
structure of the source \citep{bjor-blu}. Curvature in the spectrum 
of electrons or the superposition of two or more independent
components with different spectral indices would also naturally lead to FDP
\citep{b18}.

FDP can be manifested in relation to both the polarisation degree and
the polarisation vector (FDPA), but our dataset contains little indication of
the latter. An appreciable level of FDPA ($\theta_{I}-\theta_{V}
\lesssim 5^\circ$) is only seen during the first
and second nights of observations (see Table A1),
after which it gradually vanishes as the emission of the variable
component increases towards the end of the campaign and starts to
dominate the contribution to the polarised emission. The polarisation 
degree $p$ has nevertheless shown
significant dependence on the observing frequency, and a trend of
increasing polarisation with frequency is apparent when the
source is at a high polarisation state (see Figure 4). The
magnitude of the observed FDP, measured as $p_{\rm{V}}/p_{\rm{I}}$, 
varied from 0.8 at low polarisation levels to 1.1 when the
polarisation was the highest. This trend in FDP has been observed
before for this source and the $p(\nu)-p$ dependency was discussed
in detail by \cite{holmes}. 

No significant intranight variations in the FDP are observed, in
connection with changes in the spectral index. This can be understood
from the fact that the flux of the extended component, with roughly
constant polarisation and spectral properties, is dominant, and
therefore masks the intranight changes which would be induced in
association with the flux microvariability. In fact, only for the
third night, where the amplitude of the intranight variations were
largest, have we seen any significant signature for intranight
changes in the degree of FDP. On the other hand, when 
the longer-term increase in the photometric flux of the source is
combined with an increase on the intrinsic polarisation degree of the
variable component, as seem towards the final nights of the campaign,
the dependency becomes noticiable.      

The data presented in Figure 4 are corrected for the host galaxy's
contribution according to \cite{dominici}, and it is clear that a
constant source of unpolarised emission such as the red stellar
continuum cannot account for the observed time-variability of the FDP 
\citep{b22b, b22c}. A similar argument can be invoked to rule
out contributions from thermal accretion disc emission, whose effect
would be to dilute the observed blue trend \citep{b22a}. These
arguments point to a FDP that is intrinsic to the synchrotron
source. 

In this case, a positive FDP, associated with an increase in
the polarisation degree and optical flux, can be directly associated with
the temporal evolution of a growing shock in the jet as discussed by 
\cite{val}. In their model, a shock is responsible for the
production of highly polarised radiation with a flat-spectrum
distribution that will appear superposed on the low-level polarised
emission from the extended jet, which has a steeper spectrum 
(i.e. an older particle population). The newly developed
shock will therefore introduce an
excess of high-frequency radiation from freshly accelerated particles
which, being more polarised than the extended component, will lead to
a strong FDP towards the blue, coinciding with a maximum in both
flux and polarisation degree. As the shock-accelerated electrons cool,
the flux decreases and the spectrum of the shocked component steepens,
causing the excess contribution of the highly polarised synchrotron
component to shift towards the red, supressing or changing the sign
of the FDP that now is greater towards the red. Figure 4 shows this
trend very clearly, as we observe FDP increasing towards the blue during the
high states which inverts towards the red when the polarised flux is
minimum.

\begin{figure}
\begin{center}
\rotatebox{-90}{\resizebox{7.2cm}{8.5cm}{\includegraphics{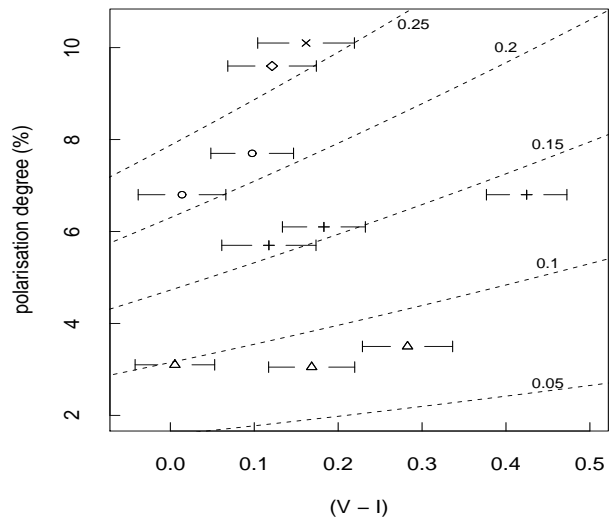}}}
 \caption{Plot of the V-band polarisation degree versus V-I
   spectral index. The dashed lines indicate positions of constant degree of
   ordering of the magnetic field $B_{o}/B_{c}$, as indicated by the
   labels. Same symbols refer to data taken during a specific night:
   day 1 ($\circ$), day 2 ($\triangle$), day 3 (+), day 4 ($\times$)
   and day 5 ($\diamond$). No multi-band observations are available
   for the last night of the campaign.} 
\end{center}
\label{bfield}
\end{figure}

\subsection{Timescales of Magnetic Field Evolution}

\cite{bjor} suggests that multiple-component models can be regarded as
an approximation to what is in reality a more complex synchrotron
source whose properties vary from one point to another, and in which
one or more components dominate the emission at given epochs.
More insight into the structure of the source's magnetic field
structure can then be obtained following an argument by \cite{ks}. 
Changes in the degree of polarisation $p$ of a synchrotron
source are directly related to the evolution of the magnetic field
structure in the emitting region, which consists of the superposition
of an ordered ($B_{o}$, provided by the shock) plus a chaotic magnetic
field component ($B_{c}$). Those authors show that the magnitude of
$p$ at any given time depends only on the spectral index of the
emission $\alpha = (\gamma-1)/2$ and the amount of field ordering 
$\beta = B_{o}/B_{c}$. At the limit of small $\beta$, we have:

\begin{equation}
p = f(\gamma)\beta^2=\frac{(\gamma+3)(\gamma+5)}{32}~\Pi_{0}~\beta^2
\label{eq01}
\end{equation}

Here $f(\gamma)$ is a slowly varying function of $\alpha$
\citep{sazo}, and $\Pi_{0} = (\gamma+1)/(\gamma+7/3)$ is the
polarisation degree of a perfectly uniform magnetic field. The
observed range of spectral indices, resulting from the acceleration
and cooling of particles in the variable shock component, imply only a 
narrow range for $f(\gamma)$ (0.5 - 0.8), which is in itself
insufficient to explain the entire range of variations observed in $p$.
This means that significant internight changes of the degree of field 
ordering must also be present to account for the observed polarisation 
variability, as also expected from the discussion of an inhomogeneous jet
in Section 4.1. 

The variation in the degree of ordering of the field is shown in
Figure 5, where the dashed lines correspond to different fractions of 
$B_{o}/B_{c}$, calculated from equation (5). During our observations, $\beta$
varied between 10-25\%. Figure 5 also shows that values for $\alpha$ 
corresponding to a same night tend to align along the directions of
constant $\beta$ indicating that changes in the spectral index happen on
shorter timescales than those of the magnetic field and that therefore
the timescales for particle cooling and acceleration are decoupled
from those of changes in $B$. If the ordering of the field is
provided by shock compression, the relative amount of ordering can be
related to the shock strenght at a given instant. In this sense, one
can notice that the increase in flux level seen towards the end of the 
campaign correlates with the two nights with higher $B_{0}/B_{c}$ in Figure 5.

\section{Conclusions}

In this paper we presented optical polarimetric observations of 
PKS 2155-304 observed in quiescence. Supported by correlated
optical and radio VLBI polarisation properties, we conducted a detailed
analysis of the source's variability. We showed that its emission
properties are consistent with the optical polarised flux having
origin at the polarised radio core. The structure of the source is
modeled as an inhomogeneous synchrotron source consisting of an
underlying jet with tangled field which is locally ordered by shock
compression of the flow. 

It is a common feature of BL Lacs that the flux and polarisation
variations show no obvious temporal correlations. We have
analysed the possible sources of variability within a shock-in-jet
model and have concluded that the flux micro-variability can be
interpreted as direct signature of particle acceleration and cooling
at the shock front. We have also observed variations of the spectral
index on timescales of a few hours which support this picture. 
The longer timescales of the polarisation variability are associated
with the propagation of the shock along a structured jet with
changing physical properties. This picture was suggested to be linked
with results of general-relativistic MHD simulations by Mckinney, which
predict the formation of instability-induced ``patches'' in the jet at 
sub-parsec scales.

Our model for the optical emission shows that most of the optical flux
originates in the weakly polarised, stable jet component, whereas the
photo-polarimetric variability results from the development
and propagation of a shock in the jet. We used this multi-zone jet
scenario to support the idea proposed by \cite{b1c} that a multi-zone
scenario is necessary to describe the temporal beahviour of BL Lacs in
the quiescent state.
Whereas most of the optical flux has its origin in the extended jet
component, the variable optical emission seems to originate
in a shock component, with higher Doppler factors and magnetic field
intensities than modeled by \cite{kat} and \cite{b1c} for the extended
jet. A consequence of this scenario is that the optical
polarimetric emission is potentially a better tracer of the high-energy 
emission, revealing the importance of  optical polarimetric monitoring in
multiwavelenght campaigns. 

In fact, if the variable and polarised optical and TeV emissions are indeed
associated, then the radio core could be identified as the
source of the quiescent TeV flux, but this hypothesis must
be thoroughly tested with further simultaneous observations of this and other
objects. Furthermore, in the case this association holds, the IC flux
would be correlated rather with the behaviour of the variable shock component,
responsible for the polarimetric variability than the extended jet
component. This fact also needs to be tested, preferentially via
multiwavelenght and optical polarimetric observations of the source
conducted during a high state.

Polarimetric observations are an essential ingredient of MWL 
campaigns if one wishes to put strong constraints to the site and
physical properties of the emission. The present paper is the first
as part of a long-term project under development at the National
Astrophysics Laboratorory (LNA/Brazil) to study the optical
polarimetric properties of TeV blazars. Results of other campaigns
will be presented in due course.

\section*{Acknowledgments}

The authors thank Profs. M. Birkinshaw and H. Marshall for
useful comments and discussions on the analysis of the data presented
here. U. Barres de Almeida acknowledges a Ph.D. Scholarship from the
CAPES Foundation, Ministry of Education of
Brazil. G.A.P. Franco and Z. Abraham acknowledge CNPq for partially
supporting their research.

%\appendix
%\section{}

\onecolumn
\begin{longtable}{|ccccc|}
\caption[]{Journal of LNA polarimetric observations \label{tab:data}}\\
\endfirsthead
\caption[]{continued}\\
\hline
 MJD    & Filter & Flux  & P    & P.A.\\
 (-54000)&        & (mJy) & (\%) & $(^\circ)$ \\
\hline
\endhead
%\hline \hline \\[1.3ex]
%\begin{center}
%\small{
%\begin{longtable}{|cccccc|}
\hline
  MJD    & Filter & Flux  & P    & P.A.\\
 (-54000)&        & (mJy) & (\%) & $(^\circ)$ \\
\hline
\multicolumn{5}{|c|}{Sep.01}\\
  712.54 & V & 27.550 (.011) & 6.73 (.06) & 89.0 (0.2)\\
  712.57 & R & 28.748 (.011) & 6.36 (.05) & 88.7 (0.2)\\
  712.58 & I & 31.458 (.013) & 5.96 (.03) & 86.0 (0.1)\\ 
  712.61 & R & 27.105 (.011) & 5.86 (.05) & 90.1 (0.2)\\
  712.63 & R & 27.411 (.011) & 5.76 (.05) & 90.0 (0.2)\\
  712.64 & R & 26.313 (.014) & 5.81 (.09) & 90.2 (0.4)\\
  712.65 & R & 27.366 (.014) & 5.76 (.08) & 90.1 (0.3)\\
  712.66 & R & 27.882 (.011) & 5.90 (.08) & 89.9 (0.4)\\
  712.67 & R & 27.325 (.011) & 5.75 (.04) & 89.3 (0.1)\\
  712.67 & V & 27.736 (.011) & 5.85 (.02) & 91.4 (0.1)\\
  712.69 & R & 27.947 (.014) & 5.65 (.02) & 89.8 (0.1)\\
  712.69 & I & 30.864 (.014) & 5.55 (.06) & 85.6 (0.3)\\
  712.71 & R & 27.678 (.020) & 5.70 (.05) & 88.9 (0.2)\\
  712.72 & R & 26.723 (.081) & 5.63 (.06) & 88.3 (0.3)\\
  712.73 & R & 27.117 (.054) & 5.58 (.07) & 87.8 (0.3)\\
  712.74 & R & 26.615 (.088) & 5.41 (.05) & 87.7 (0.2)\\ 
  712.75 & R & 25.716 (.065) & 5.41 (.07) & 87.6 (0.4)\\
\hline
\multicolumn{5}{|c|}{Sep.02}\\
  713.48 & V & 25.017 (.013) & 2.59 (.05) & 95.0 (0.5)\\
  713.49 & R & 25.688 (.021) & 2.64 (.04) & 95.5 (0.4)\\
  713.50 & I & 29.440 (.030) & 2.77 (.04) & 91.0 (0.4)\\
  713.52 & R & 26.016 (.022) & 2.88 (.07) & 96.7 (0.7)\\
  713.53 & R & 25.448 (.137) & 2.67 (.13) & 94.9 (1.4)\\
  713.54 & R & 24.946 (.011) & 2.60 (.08) & 95.7 (0.9)\\
  713.55 & R & 25.377 (.026) & 2.69 (.10) & 96.0 (1.1)\\
  713.56 & R & 24.665 (.116) & 2.78 (.11) & 94.8 (1.1)\\
  713.57 & R & 23.993 (.010) & 2.92 (.06) & 96.9 (0.6)\\
  713.58 & R & 25.723 (.010) & 2.67 (.05) & 96.2 (0.5)\\
  713.59 & V & 25.158 (.010) & 2.64 (.05) & 97.5 (0.5)\\
  713.60 & R & 26.079 (.010) & 2.65 (.06) & 99.1 (0.6)\\
  713.61 & I & 28.203 (.015) & 2.64 (.06) & 93.1 (0.7)\\
  713.63 & R & 25.427 (.012) & 2.59 (.03) & 96.2 (0.3)\\
  713.64 & R & 25.608 (.010) & 2.48 (.05) & 98.3 (0.5)\\
  713.65 & R & 25.697 (.011) & 2.55 (.05) & 97.8 (0.5)\\
  713.66 & R & 25.247 (.010) & 2.69 (.03) & 98.1 (0.3)\\
  713.67 & R & 25.436 (.013) & 2.70 (.05) & 98.3 (0.5)\\
  713.67 & R & 25.457 (.012) & 2.74 (.02) & 98.9 (0.3)\\
  713.68 & R & 26.637 (.010) & 2.68 (.05) & 98.6 (0.5)\\
  713.69 & R & 25.567 (.011) & 2.77 (.04) & 99.8 (0.5)\\
  713.70 & R & 24.992 (.010) & 2.81 (.03) & 100.7 (0.3)\\
  713.71 & V & 24.338 (.011) & 2.95 (.04) & 99.5 (0.4)\\
  713.72 & R & 24.960 (.013) & 2.88 (.01) & 101.2 (0.1)\\
  713.73 & I & 29.701 (.012) & 3.00 (.09) & 98.8 (0.8)\\
  713.74 & R & 25.352 (.011) & 3.02 (.03) & 102.1 (0.3)\\
  713.75 & R & 25.235 (.021) & 3.02 (.03) & 102.2 (0.3)\\
\hline
\multicolumn{5}{|c|}{Sep.03}\\
  714.46 & V & 26.182 (.047) & 4.80 (.02) & 108.1 (1.5)\\
  714.47 & R & 25.648 (.015) & 4.78 (.05) & 107.3 (0.3)\\
  714.49 & I & 30.217 (.014) & 4.25 (.02) & 107.0 (1.7)\\
  714.51 & R & 25.873 (.012) & 4.66 (.06) & 108.3 (0.3)\\
  714.53 & R & 24.405 (.016) & 4.76 (.05) & 108.4 (0.3)\\
  714.54 & R & 26.439 (.012) & 4.62 (.05) & 109.2 (0.3)\\
  714.55 & R & 25.636 (.011) & 4.65 (.02) & 109.1 (0.1)\\
  714.56 & R & 26.000 (.012) & 4.63 (.04) & 109.0 (0.2)\\
  714.58 & R & 25.420 (.025) & 4.72 (.03) & 108.9 (0.2)\\
  714.59 & R & 31.513 (.030) & 4.76 (.06) & 109.3 (0.4)\\
  714.60 & R & 27.238 (.011) & 4.83 (.04) & 109.1 (0.2)\\
  714.61 & V & 25.163 (.014) & 5.17 (.04) & 109.9 (0.2)\\
  714.63 & R & 32.653 (.015) & 5.31 (.06) & 113.4 (0.3)\\
  714.64 & I & 29.723 (.016) & 4.96 (.06) & 109.7 (0.3)\\
  714.65 & R & 25.942 (.010) & 5.24 (.05) & 110.1 (0.3)\\
  714.66 & R & 26.953 (.011) & 5.29 (.04) & 109.5 (0.2)\\
  714.67 & R & 26.921 (.018) & 5.40 (.05) & 109.7 (0.2)\\
  714.68 & R & 28.521 (.022) & 5.33 (.09) & 109.9 (0.4)\\
  714.69 & R & 24.525 (.022) & 5.46 (.05) & 109.4 (0.3)\\
  714.70 & R & 25.589 (.016) & 5.43 (.03) & 110.3 (0.1)\\
  714.71 & R & 27.348 (.016) & 5.48 (.02) & 110.3 (0.1)\\
  714.72 & R & 26.461 (.018) & 5.42 (.03) & 110.5 (0.1)\\
  714.73 & V & 23.475 (.021) & 5.71 (.10) & 111.4 (0.5)\\
  714.75 & R & 24.800 (.059) & 5.65 (.05) & 109.9 (0.2)\\
  714.75 & I & 29.973 (.064) & 5.44 (.07) & 109.2 (0.3)\\
\hline
\multicolumn{5}{|c|}{Sep.04}\\
  715.56 & R & 26.323 (.073) & 8.24 (.10) & 114.6 (0.3)\\
  715.57 & R & 25.697 (.031) & 8.26 (.06) & 115.1 (0.2)\\
  715.58 & R & 26.340 (.011) & 8.26 (.06) & 114.7 (0.2)\\
  715.59 & R & 26.004 (.054) & 8.27 (.07) & 114.8 (0.2)\\
  715.60 & R & 26.566 (.019) & 8.15 (.02) & 115.1 (0.1)\\
  715.61 & R & 26.088 (.035) & 8.45 (.06) & 114.6 (0.2)\\
  715.62 & R & 25.598 (.011) & 8.19 (.05) & 115.0 (0.1)\\
  715.63 & V & 26.745 (.078) & 8.56 (.02) & 115.7 (0.1)\\
  715.64 & R & 26.366 (.011) & 8.17 (.06) & 114.7 (0.2)\\
  715.65 & I & 31.219 (.013) & 7.60 (.08) & 114.9 (0.2)\\
  715.66 & R & 26.293 (.010) & 8.21 (.04) & 114.6 (0.1)\\
  715.67 & R & 26.603 (.013) & 8.12 (.04) & 114.9 (0.1)\\
  715.68 & R & 26.427 (.010) & 8.06 (.02) & 114.3 (0.1)\\
  715.70 & R & 25.667 (.010) & 8.04 (.03) & 114.5 (0.1)\\
  715.71 & R & 26.795 (.011) & 8.03 (.05) & 114.8 (0.1)\\
  715.72 & R & 25.930 (.011) & 7.96 (.07) & 114.6 (0.2)\\
\hline
\multicolumn{5}{|c|}{Sep.05}\\
  716.62 & R & 27.500 (.011) & 7.76 (.02) & 116.2 (0.8)\\
  716.63 & R & 27.190 (.012) & 7.79 (.14) & 116.3 (0.5)\\
  716.65 & R & 23.736 (.032) & 8.24 (.16) & 116.3 (0.5)\\
  716.66 & R & 25.617 (.013) & 7.79 (.21) & 116.3 (0.7)\\
  716.67 & R & 24.193 (.018) & 7.87 (.17) & 116.9 (0.6)\\
  716.68 & R & 28.050 (.012) & 7.81 (.21) & 116.8 (0.7)\\
  716.68 & V & 26.767 (.011) & 8.18 (.01) & 119.9 (0.6)\\
  716.70 & R & 28.764 (.011) & 7.66 (.18) & 118.1 (0.6)\\
  716.70 & I & 30.864 (.012) & 7.12 (.06) & 119.0 (0.2)\\
  716.72 & R & 27.295 (.012) & 7.92 (.17) & 116.6 (0.6)\\
\hline
\multicolumn{5}{|c|}{Sep.06}\\ 
  717.62 & R & 30.691 (.098) & 5.80 (.09) & 130.6 (0.4)\\
  717.65 & R & 31.227 (.174) & 5.79 (.40) & 131.6 (1.9)\\
  717.66 & R & 31.331 (.168) & 5.62 (.07) & 131.1 (0.4)\\
  717.68 & R & 30.747 (.063) & 5.55 (.05) & 131.8 (0.2)\\
  717.69 & R & 31.253 (.085) & 5.58 (.14) & 132.0 (0.7)\\
  717.70 & R & 30.847 (.243) & 5.36 (.07) & 132.2 (0.4)\\
  717.71 & R & 32.223 (.239) & 5.28 (.08) & 132.1 (0.4)\\
\hline
%\end{longtable}
%\end{center}
\end{longtable}
\twocolumn
%\bsp

\label{lastpage}


\begin{thebibliography}{99}

\bibitem[\protect\citeauthoryear{Aharonian et al.}{Aharonian et al.}{2005}]{b0c} Aharonian F. et al., 2005, A\&A, 442, 895
\bibitem[\protect\citeauthoryear{Aharonian et al.}{Aharonian et al.}{2009}]{b1c} Aharonian F. et al., 2009, ApJ, 696, L150
\bibitem[\protect\citeauthoryear{Aharonian et al.}{Aharonian et al.}{2010}]{prep} Aharonian F. et al., 2010, arXiv:1005.3702
\bibitem[\protect\citeauthoryear{Allen et al.}{Allen et al.}{1993}]{allen} Allen R.G., Smith P.S. et al. 1993, ApJ, 403, 610
\bibitem[\protect\citeauthoryear{Andruchow et al.}{Andruchow et al.}{2005}]{micro} Andruchow I., Romero G. \& Celone S. 2005, A\&A, 442, 97
\bibitem[\protect\citeauthoryear{Angel et al.}{Angel \& Stockman}{1980}]{b0} Angel J.R.P. \& Sotckman H.S., 1980, ARA\&A, 18, 321
\bibitem[\protect\citeauthoryear{Bj\"{o}rnsson}{Bj\"{o}rnsson \& Blumenthal}{1982}]{bjor-blu} Bj\"{o}rnsson C.-I. \& Blumenthal G.R. 1982, ApJ, 259, 805.
\bibitem[\protect\citeauthoryear{Bj\"{o}rnsson}{Bj\"{o}rnsson}{1985}]{bjor} Bj\"{o}rnsson C.-I. 1985, MNRAS, 216, 241.
\bibitem[\protect\citeauthoryear{Brindle et al.}{Brindle et al.}{1996}]{brindle} Brindle C., 1996, MNRAS, 282, 788
\bibitem[\protect\citeauthoryear{Courvoisier et al.}{Courvoisier et al.}{1995}]{cour} Courvoisier T.J.-L., Blecha A. et al. 1995, ApJ, 438, 108.
\bibitem[\protect\citeauthoryear{Dominici et al.}{Dominici et al.}{2006}]{dominici} Dominici T., Abraham Z. \& Galo, A. 2006, A\&A, 460, 665.
\bibitem[\protect\citeauthoryear{Fermi/LAT Collab.}{Fermi/LAT Collab.}{2010}]{Fermib} Fermi/LAT Collab. et al. 2010, arXiv:1004.2857
\bibitem[\protect\citeauthoryear{Gabuzda et al.}{Gabuzda et al.}{2006}]{gabuzda} Gabuzda D.C. et al. 2006, MNRAS, 369, 1596.
\bibitem[\protect\citeauthoryear{Giebels et al.}{Giebels et al.}{2002}]{b5} Giebels B., Bloom E.D., et al., 2002, ApJ, 571, 763.
\bibitem[\protect\citeauthoryear{Ginzburg}{Ginzburg \& Syrovatskii}{1965}]{b6} Ginzburg V.L. \& Syrovatskii S.I., 1965, ARA\&A, 3, 297.
\bibitem[\protect\citeauthoryear{Hagen-Thorn, et al.}{Hagen-Thorn et al.}{2008}]{b11} Hagen-Thorn V.A. et al., 2008, AJ, 672, 40
\bibitem[\protect\citeauthoryear{Holmes et al.}{Holmes et al.}{1984}]{holmes} Holmes V.A., Brand W.J.L. et al., 1984, MNRAS, 211, 497
\bibitem[\protect\citeauthoryear{Hughes et al.}{Hughes et al.}{1989}]{hughes} Hughes P.A., Aller H.D.\& Aller M.F. 1989, ApJ, 341, 68.
\bibitem[\protect\citeauthoryear{Jones et al.}{Jones et al.}{1985}]{jones} Jones T.W., Rudnik L., et al. 1985, AJ, 290, 627
\bibitem[\protect\citeauthoryear{Jones}{Jones}{1988}]{jones88} Jones T.W. 1988, AJ, 332, 678
\bibitem[\protect\citeauthoryear{Jorstad et al.}{Jorstad et al.}{2001}]{jorstad01} Jorstad S.G., Marscher A.P. et al. 2001, ApJ, 556, 738.
\bibitem[\protect\citeauthoryear{Jorstad et al.}{Jorstad et al.}{2007}]{jorstad07} Jorstad S.G., Marscher A.P. et al. 2007, AJ, 134, 799.
\bibitem[\protect\citeauthoryear{Katarzy\'{n}ski}{Katarzy\'{n}ski et al.}{2008}]{kat} Katarzy\'{n}ski K., Lenain J.P. et al., 2008, MNRAS, 390, 371
\bibitem[\protect\citeauthoryear{Korchakov}{Korchakov \& Syrovatskii}{1962}]{ks} Korchakov A.A. \& Syrovatskii S.I., 1962, Sov. Astr., 5, 5.
\bibitem[\protect\citeauthoryear{Laing}{Laing}{1980}]{laing} Laing R.A. 1980, MNRAS, 193, 439.
\bibitem[\protect\citeauthoryear{Lister et al.}{Lister \& Smith}{2000}]{b15} Lister M.L. \& Smith P.S. 2000, ApJ, 541, 66.
\bibitem[\protect\citeauthoryear{Marshall et al.}{Marshall et al.}{2002}]{marshall} Marshall H.L., Miller B.P., Davis D.S. et al. 2002, ApJ, 564, 683.
\bibitem[\protect\citeauthoryear{Marscher et al.}{Marscher \& Gear}{1985}]{shock} Marscher A.P. \& Gear W.K. 1985, ApJ, 298, 114.
\bibitem[\protect\citeauthoryear{McKinney}{McKinney}{2005a}]{mckinneya} McKinney J.C., 2005a, astro.ph..6368
\bibitem[\protect\citeauthoryear{McKinney}{McKinney}{2005b}]{mckinneyb} McKinney J.C., 2005b, astro.ph..6369
\bibitem[\protect\citeauthoryear{Moore et al.}{Moore et al.}{1982}]{moore} Moore R.L., McGraw J.T., Angel J.R.P. et al.1982, ApJ, 260, 415.
\bibitem[\protect\citeauthoryear{Nordsieck}{Nordsieck}{1976}]{b18} Nordsieck K.H., 1976, ApJ, 209, 653
\bibitem[\protect\citeauthoryear{Pacholczyk}{Pacholczyk}{1970}]{pac} Pacholczyk A.G., 1970, Radio Astrophysics, Freeman, San Francisco.
\bibitem[\protect\citeauthoryear{Pereyra}{Pereyra}{2000}]{b19} Pereyra A., 2000, Ph.D. Thesis, University of S\~{a}o Paulo
\bibitem[\protect\citeauthoryear{Piner et al.}{Piner \& Edwards}{2004}]{b20} Piner B.G. \& Edwards P.G., 2004, ApJ, 600, 115.
\bibitem[\protect\citeauthoryear{Piner et al.}{Piner et al.}{2008}]{b21} Piner B.G., Pant N. \& Edwards P.G., 2008, ApJ, 678, 64.
\bibitem[\protect\citeauthoryear{Qian}{Qian et al.}{1991}]{qian} Qian S.J., Quirrenbach A. et al. 1991, A\&A, 241, 15.
\bibitem[\protect\citeauthoryear{Qian}{Qian}{1993}]{qian93} Qian S.J., 1993, CA\&A, 17, 229.
\bibitem[\protect\citeauthoryear{Rector \& Perlman}{Rector \& Perlman}{2003}]{b22rp} Rector T.A. \& Perlman E.S. 2003, AJ, 126, 47.
\bibitem[\protect\citeauthoryear{Rees}{Rees}{1967}]{rees} Rees M.J., 1967, MNRAS, 135, 345
\bibitem[\protect\citeauthoryear{Saikia}{Saikia \& Salter}{1988}]{saik} Saikia D.J. \& Salter C.J., 1988, ARA\&A, 26, 93.
\bibitem[\protect\citeauthoryear{Sazonov}{Sazonov}{1972}]{sazo} Sazonov V.N., 1972, Ap\&SS, 19, 25.
\bibitem[\protect\citeauthoryear{Scargle}{Scargle}{1982}]{b22} Scargle J.D., 1982, ApJ, 263, 835.
\bibitem[\protect\citeauthoryear{Smith et al.}{Smith \& Sitko}{1991}]{b22a} Smith P.S. \& Sitko M.L., 1991, ApJ, 383, 580
\bibitem[\protect\citeauthoryear{Smith et al.}{Smith et al.}{1991}]{b22b} Smith P.S., Jannuzi B.T. \& Elston R., 1991, ApJS, 77, 67
\bibitem[\protect\citeauthoryear{Smith et al.}{Smith et al.}{1992}]{b22c} Smith P.S., Hall P.B. et al., 1992, ApJ, 400, 115
\bibitem[\protect\citeauthoryear{Tommasi et al.}{Tommasi et al.}{2001}]{tommasi} Tommasi L., D\'{i}az R., Palazzi E., et al. 2001, ApJ, 132, 73
\bibitem[\protect\citeauthoryear{Valtaoja et al.}{Valtaoja et al.}{1991}]{val} Valtaoja L., Valtaoja E. et al. 1991, AJ, 101, 78
\bibitem[\protect\citeauthoryear{Valtaoja et al.}{Valtaoja et al.}{1991b}]{valb} Valtaoja L., Valtaoja E. et al. 1991b, AJ, 102, 1946.

\end{thebibliography}
\end{document}